\documentclass{article}

\usepackage[preprint]{neurips_2024}

\usepackage[utf8]{inputenc} % allow utf-8 input
\usepackage[T1]{fontenc}    % use 8-bit T1 fonts
\usepackage{hyperref}       % hyperlinks
\usepackage{url}            % simple URL typesetting
\usepackage{booktabs}       % professional-quality tables
\usepackage{amsfonts}       % blackboard math symbols
\usepackage{nicefrac}       % compact symbols for 1/2, etc.
\usepackage{microtype}      % microtypography
\usepackage{xcolor}         % colors
\usepackage{listings}
\usepackage{comment}
\usepackage{graphicx}
\usepackage{natbib}

\title{Beyond Multiple Instance Learning: Full Resolution All-In-Memory End-To-End Pathology Slide Modeling}

\author{%
  Gabriele~Campanella \\
  Windreich Department of AI and Human Health \\
  Icahn School of Medicine at Mount Sinai \\
  New York, NY 10029 \\
  \texttt{gabriele.campanella@mssm.edu} \\
  \And
  Eugene~Fluder \\
  Scientific Computing and Data \\
  Icahn School of Medicine at Mount Sinai \\
  New York, NY 10029 \\
  \AND
  Jennifer~Zeng \\
  Department of Pathology, Molecular and Cell-Based Medicine \\
  Icahn School of Medicine at Mount Sinai \\
  New York, NY 10029 \\
  \And
  Chad~Vanderbilt \\
  Department of Pathology \\
  Memorial Sloan Kettering Cancer Center \\
  New York, NY 10065 \\
  \And
  Thomas~J.~Fuchs \\
  Windreich Department of AI and Human Health \\
  Icahn School of Medicine at Mount Sinai \\
  New York, NY 10029 \\
}

\begin{document}

\maketitle

\begin{abstract}
  Artificial Intelligence (AI) has great potential to improve health outcomes by training systems on vast digitized clinical datasets. Computational Pathology, with its massive amounts of microscopy image data and impact on diagnostics and biomarkers, is at the forefront of this development. Gigapixel pathology slides pose a unique challenge due to their enormous size and are usually divided into tens of thousands of smaller tiles for analysis. This results in a discontinuity in the machine learning process by separating the training of tile-level encoders from slide-level aggregators and the need to adopt weakly supervised learning strategies. Training models from entire pathology slides end-to-end has been largely unexplored due to its computational challenges. To overcome this problem, we propose a novel approach to jointly train both a tile encoder and a slide-aggregator fully in memory and end-to-end at high-resolution, bridging the gap between input and slide-level supervision. While computationally expensive, detailed quantitative validation shows promise for large-scale pre-training and fine-tuning of pathology foundation models.
\end{abstract}

\section{Introduction}

The application of Artificial Intelligence (AI) in the medical domain has the potential to improve health and disease outcomes of the population at large. The advent of AI in healthcare is driven by the digitization of vast quantities of clinical data. In recent years, pathology departments around the world have started transitioning to a digital workflow which includes scanning pathology slides, paving the way for the emergence of computational pathology and the development of AI-based systems for diagnosis and prognosis in pathology. Compared to other medical image modalities, digitized pathology slides are orders of magnitude larger resulting in giga-pixel images that can span over 100,000 pixels in each dimension at 40x magnification. Processing these very large images is challenging, leading to the common strategy in pathology to divide slides into tens of thousands of small tiles for analysis. Weakly supervised learning is often used, frequently formalized with multiple instance learning (MIL)~\cite{dietterich_solving_1997}, to train predictive models.

Most works in computational pathology fall into two main categories: 1) approaches that focus on training a tile-level encoder, and 2) approaches that focus on training a slide-level aggregator, commonly leveraging a pre-trained encoder for feature extraction. Tile-level encoders can be trained to directly extract relevant features for specific tasks. Traditionally this was achieved with supervised learning through the use of pixel-level annotations, but this approach suffered from the lack of annotated datasets. The introduction of weakly supervised learning marked a paradigm shift and allowed for the training of tile-encoders directly without the need for annotations. A notable example is~\cite{campanella_clinical-grade_2019} where a clinical-grade decision support system was trained on a large scale using max-pool MIL. In contrast, aggregator strategies rely on extracting features from tiles using a frozen encoder which was trained on a different task~\cite{ilse_attention-based_2018,shao_transmil_2021}. In this approach, the tile-level encoder is not optimized. Originally, ImageNet pre-trained encoders were used for this purpose. More recently, with the success of self-supervised learning (SSL), pre-trained models on pathology data have been published~\cite{kang_benchmarking_2022,wang_transformer-based_2022,filiot_scaling_2023,chen_scaling_2022,chen_general-purpose_2023,vorontsov_virchow_2023,campanella_computational_2023,chen_towards_2024} and a transition from ImageNet based encoders to pathology tailored ones is occurring. While SSL strategies applied to pathology can achieve better performance in various downstream tasks compared to ImageNet trained models~\cite{campanella_computational_2023}, it is still unknown whether current SSL strategies result in encoders that show optimal downstream task performance.

Naively training a model end-to-end on entire pathology slides at high resolution would be too memory intensive for a single GPU and not a lot of research has been devoted to this line of work. End-to-end training requires that the loss generated at the slide-level is backpropagated all the way to the image pixels.
\cite{qu_bi-directional_2022}, proposed to train both a tile encoder and slide aggregator by using the attention scores of the aggregator to guide tile selection for training the encoder in teacher-student framework. While both encoder and aggregator are trained, this work can't be considered end-to-end since there is a disconnect between the two training stages.
\cite{xie_beyond_2020} presented a method that allows to train an encoder directly from slide labels by clustering the feature space and representing a slide as a list of tiles closest in feature space to learned prototypes. While this method performs end-to-end training, it is restricted to a small subset of tiles and can't be extended to analyze pixel data from an entire slide.

Pioneering work from~\cite{pinckaers_streaming_2022} proposed for the first time a method to train a convolutional neural network (CNN) end-to-end from pixels to slide labels on images at high resolution. They introduced ``streaming convolutions'' where they leverage an ad-hoc check-pointing strategy, trading off compute for memory consumption, to perform a stack of 2D-convolution operations on images of arbitrarily large size. While their original proof of concept was tested on down-sampled slides of about 8k pixels, the authors improved the method in~\cite{pinckaers_detection_2021,dooper_gigapixel_2023}, eventually analyzing up to 65k pixel images, where a majority of pathology slides can be analyzed at high resolution. Interestingly, in REF they also introduced an aggregator on top of the CNN-based backbone which aligns their efforts with the most recent trends in computational pathology literature. We find that their latest method proposed in REF, presents the following drawbacks: 1) Due to the need to recompute intermediate activations, streaming convolutions are several times slower than vanilla convolutions. 2) The method relies on the locality of convolutional operations. Due to this, the portions of the model that are streamed can't contain global operations such as batch-normalization and self-attention. More generally, it is not possible to leverage any arbitrary encoder network, including the now popular transformer-based pathology foundation models (\cite{chen_general-purpose_2023,campanella_computational_2023,vorontsov_virchow_2023,chen_towards_2024}), limiting the potential of this method in the fine-tuning setting. 3) The most advanced version of this approach uses a ResNet34 as the CNN backbone, which may limit the representation capacity and downstream performance due to its limited size. While larger models could be used, they become impractical with current hardware.

More recently,~\cite{wang_when_2023} introduced LongViT where they applied LongNet~\cite{ding_longnet_2023}, a transformer model that uses a self-attention approximation with linear scaling memory, to computational pathology. They are able to train a vision transformer (ViT) from image pixels over an entire slide by relying on a sparse self-attention mechanism and GPU parallelization. This line of work is promising but in its early stages and currently suffers from several drawbacks: 1) Their pipeline relies on drastically down-sampling the slide resolution, potentially losing cellular level information which may be important for certain tasks. Currently they tested its use with images of up to 32k pixels, still relatively small for pathology data. 2) As per ViTs, they encode very small patches (16 or 32 pixels), generating sequences of millions of tokens, and rely on self-attention to obtain global representations. As such, most of the information is encoded in the spatial relationship between the small visual tokens. Modeling sequences of millions of tokens is challenging. 3) It is still to be seen whether LongNet, which was designed for 1D signals, is the right choice for approximating self-attention for 2D signals. In particular, LongViT's concept of locality applies to the flattened sequence of tokens. Hence, neighbor tokens vertically are not considered near using LongNet's sparse attention. 4) Their framework is less suitable for leveraging pre-trained foundation models.

In contrast to previous efforts in this space, namely streaming convolutions and LongViT, which rely on a single slide-level network, we follow the popular two step strategy of leveraging a deep tile-level encoder and a global aggregator. We propose to optimize both fully in memory and end-to-end at high-resolution, where the slide-level loss is backpropagated directly through the aggregator and tile encoder. We achieve this via GPU-parallelization and customized GPU-GPU communication.
Compared to the other methods, the strengths of this approach are: 1) High speed. We require just a single pass through the network, in contrast to the checkpointing strategy in streaming convolutions. 2) Flexibility in architecture design. Any combination of encoder and aggregator can be supported. 3) Can accommodate the use of pre-trained foundation models for fine-tuning. 4) While entire slides can be processed at once with enough resources, it is straightforward to reduce the number of tiles analyzed at each pass, enabling the use of this framework in resource constrained settings.

The rest of the manuscript is structured as follows. First, we explain the method in detail, providing pseudo-code and a pytorch implementation. We prove the equivalence of the proposed multi-GPU framework with its single GPU counterpart. Then, we demonstrate the effectiveness of our method on a clinically relevant task in cancer research, EGFR mutation prediction in lung adenocarcinoma (LUAD), for which we have thoroughly investigated other training strategies, including fully supervised and weakly supervised strategies, tile-level and slide-level algorithms, as well as the use of large scale self-supervised learning pre-training. Next, we show that it is possible to optimize encoder/aggregator models on entire pathology slides at high resolution by training a breast cancer detector. Finally, we use this framework to fine-tune a ViT-base pathology foundation model for the task of predicting EGFR mutational status in LUAD patients, achieving superior performance compared to current foundation model-based strategies.

\section{Method}

\begin{figure}
\includegraphics[width=\textwidth]{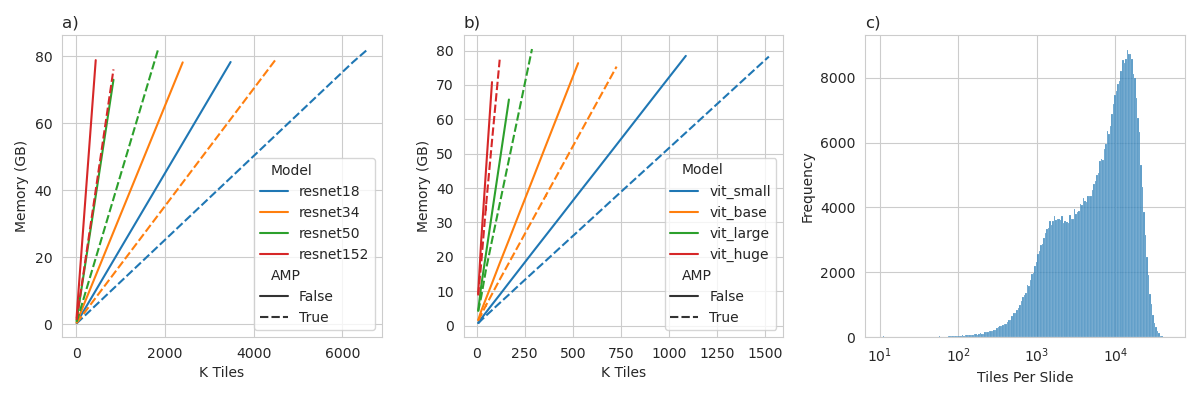}
\caption{a) H100 GPU memory usage for training ResNet encoders of different sizes. b) H100 GPU memory usage for training ViT encoders of different sizes. Full precision and AMP training are compared. c) Distribution of 20x magnification non-overlapping tissue tiles per slide in a large health system-level dataset.} \label{fig1}
\end{figure}

A modern GPU allows to jointly train a tile encoder and slide aggregator end-to-end by sampling $K$ tiles from each slide in each optimization step. This can be considered an extension of MIL where, from the original bags, $K$-sized pseudo-bags are sampled instead. This simple strategy can be improved by choosing which tiles to sample based on more complex criteria. For example, in~\cite{xie_beyond_2020} a memory bank of cluster centroids is used to select a set of heterogeneous tiles from each slide. We computed the the training memory requirements of the end-to-end pseudo-bag strategy for various popular vision architectures including convolutional neural networks (Fig.~\ref{fig1}a) and vision transformers (Fig.~\ref{fig1}b). With an 80GB H100 GPU, it is possible to jointly train encoder and aggregator with at most 840 tiles for a ResNet50, and 528 tiles for a ViT-base model. By using automatic mixed precision (AMP) with casting to 16-bit floats, it is possible to boost these numbers to 1,848 and 728 tiles for a ResNet50 and a ViT-base respectively.
%For these experiments we generate a dummy loss from the features extracted by the encoder without the use of a slide aggregator. We show that on an H100 GPU with 80GB of RAM, it is possible to run forward and backward passes with at most 840 tiles for a ResNet50, and 528 tiles for a ViT-base model. We can further boost $K$, and increase processing speeds, by using mixed precision techniques such as automatic mixed precision (AMP). Using AMP with automatic casting to 16-bit float precision it is possible to analyze 1,848 and 728 tiles for a ResNet50 and a ViT-base respectively.

We can parallelize feature encoding by assigning $K$ tiles to $N$ GPUs to encode $NK$ tiles. Whole slide images can contain tens of thousands and potentially even hundreds of thousands of tiles. In Fig.~\ref{fig1}c we show the distribution of non-overlapping tissue tiles (256 pixels, 0.5 microns per pixel = 20x magnification) per slide for the pre-training dataset described in~\cite{campanella_computational_2023}. This dataset encompasses pathology data at a health-system scale and can be considered a good representation of clinical pathology slides in general. In this dataset, the highest number of tiles in a slide is 50,578 where 99\% of slides contain less than 27,219 tiles.% and 90\% of slides contain less than 18,766.
With a ViT-base using AMP is possible to encode a full slide using 70 and 38 H100 GPUs respectively. It is important to note that while 20x magnification is the most commonly used resolution, for certain applications where cellular and nuclear features are important, it may be necessary to use 40x magnification which will increase by a factor of 4 the number of tiles per slide.

\begin{figure}[h]
\includegraphics[width=\textwidth]{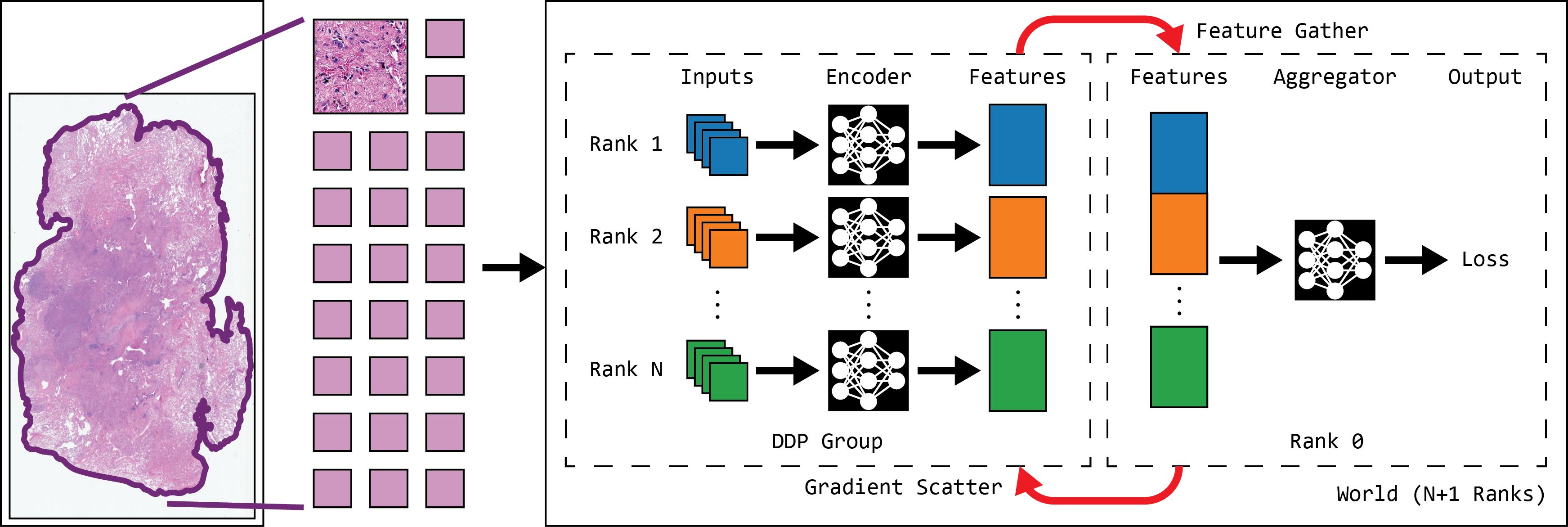}
\caption{Method overview. A distributed sampler generates appropriate batches of tissue tiles for each encoder rank in the DDP group. Features generated by the encoder ranks are gathered and concatenated in rank 0. Aggregator forward and backward passes are executed in rank 0. The gradients of the input features to the aggregator are split and scattered to the appropriate rank. In each rank a pseudo-loss is generated to continue backpropagation.} \label{fig2}
\end{figure}

To allow training of an encoder and aggregator end-to-end using the encoder parallelism described above we propose to separate encoding and aggregation to different GPU groups, where each group can consist of multiple GPUs. In our experiments, the encoding group consists of multiple GPUs, while the aggregator group is one GPU, consistent with the fact that the vast majority of current aggregation strategies rely on a single GPU. To allow for training we customize GPU-GPU communications to simulate the flow of activations and gradients as if the data was processed on a single GPU card. Assuming a GPU cluster with $N+1$ GPUs, we denote rank $0$ as the aggregator process and ranks $1,\ldots,N$ the encoder processes. For each epoch and slide, a distributed sampler directs the $N$ batches of images to the $N$ GPUs while rank $0$ waits for input. Next, the encoder processes perform a forward pass through the encoder to generate features. At this stage, feature vectors from ranks $1,\ldots,N$  are sent to rank $0$ via a gather call. The gather call breaks the computing graph and we will have to route the gradients manually during the backward pass. Rank $0$ then concatenates these into a single feature matrix that is the input to the aggregation model which generates a slide level feature vector that is projected to class logits or other output for loss computation. During the backward phase, aggregator gradients are generated from the loss $l$ in rank $0$ back to the input features. The feature gradients are then split into $N$ chunks and sent to each respective rank via a scatter call. To propagate the gradients back to the encoder model, each process with rank $1,\ldots,N$ generates a pseudo-loss $l_e$ based on the features output from the encoder $f$ and the feature gradient $g$ coming from rank $0$:
%\begin{equation}
%l_e = N \sum_{i=1}^{F} f_i g_i
%\end{equation}
\begin{math}
    l_e = N \sum_{i=1}^{F} f_i g_i
\end{math}.
We note that the pseudo-loss must be scaled by the number of processes in the encoder group $N$ to recover the expected gradient magnitude. We leverage torch’s \verb|DistributedDataParallel| (DDP) to wrap the encoder model which automatically ensures that gradients will be all-reduced and the models' weights are the same after optimization across all ranks $1,\ldots,N$. Fig.~\ref{fig2} summarizes the method.

One advantage of the proposed framework is that it can be customized to fit the needs of different use cases. The encoder architecture can be any neural network which can be initialized with custom weights if needed. The aggregator can also be chosen arbitrarily. In our experiments we focus on the popular architecture gated MIL attention~\cite{ilse_attention-based_2018}, but any aggregation method can be used. The current code base supports any aggregation as long as it can trained on a single GPU. Multi-GPU aggregators could also be supported in the future. The code will be available on GitHub while pseudo-code is provided in the Appendix in Listing~\ref{code}.
%\href{https://github.com/fuchs-lab-public/OPAL/tree/main/SEA}{GitHub}

\section{Results}

\subsection{Gradient Equivalence}
\begin{figure}
\includegraphics[width=\textwidth]{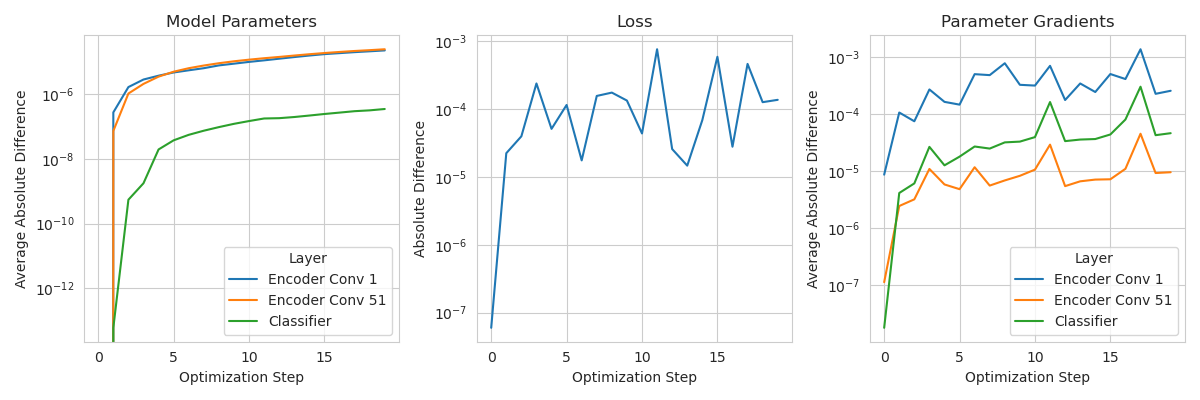}
\caption{Gradient equivalence in real world networks. We tracked parameters and gradients of three layers in the network, including two convolutional layers from the encoder and the classification layer of the aggregator. Left) Normalized L1 norm of the difference between model parameters in single and multi-GPU experiments. Middle) Absolute difference in the loss between single and multi-GPU experiments. Right) Normalized L1 norm of the difference between model parameters’ gradients in single and multi-GPU experiments.} \label{fig3}
\end{figure}

In this section we provide experimental evidence of the equivalence of single and multi-GPU runs. We inspect gradients in a simple deterministic toy network and in a non-deterministic convolutional neural network.
For the deterministic experiment, we manually set the random seed and used the following settings in torch which allow to run in a deterministic fashion: \verb|cudnn.benchmark = False| and \verb|cudnn.deterministic = True|. Both encoder and aggregator networks were composed of linear layers and ReLU non-linearities. We compared a single GPU run with an equivalent 5GPU run and established the equivalence of parameter gradients.
%The input to both runs was a randomly generated 800 by 2,048 matrix. In the single GPU run the entire input was fed through the encoder and aggregator networks, while in the parallelized version the input was split across 4 GPUs and the aggregation was done on a fifth GPU. Gradients for both networks were compared, allowing us to establish their equivalence.
Neural networks used in practice are implemented in a non-deterministic way due to the nature of GPU computations.
We tested the parallelization using a ResNet50 encoder and GMA aggregator. We stored parameters and gradients of the first convolutional layer (Encoder Conv 1), a late convolutional layer (Encoder Conv 51), and the classification layer in the aggregator (Classifier), as well as the loss at each step. To minimize stochasticity, image augmentations were not performed, GMA was instantiated without dropout, and the batch normalization layers were synchronized. We measured the normalized L1 distance between single GPU and multi-GPU runs for several optimization steps. In Fig.~\ref{fig3} we show the results of this experiment. We can see small magnitude differences in parameters, loss, and gradients which seem to plateau after a few optimization steps. In practice these differences do not have a significant effect on training as we will show in the next section.

%Here we test the framework using realistic encoders and aggregators, namely a ResNet50 and GMA. For this experiment we stored parameters and gradients of the first convolutional layer (Encoder Conv 1) and the first convolutional layer of the last block in the last layer of the encoder (Encoder Conv 51) as well as the classification layer in the aggregator (Classifier). Additionally, we also record the loss at each step. To maximize reproducibility, image augmentations were not performed, GMA was instantiated without dropout, and the batch normalization layers were synchronized.
%This is in contrast with running experiments in practice where we perform augmentations, use dropout, and do not synchronize the batch normalization layers. 
%We measured the normalized L1 distance between parameter and gradient vector of the single GPU run and the multi-GPU run and followed these measurements for several optimization steps. In Fig.~\ref{fig3} we show the results of this experiment. We can see that despite our efforts, differences in parameters, loss, and gradients can be observed. These differences have a small magnitude and seem to plateau after a few optimization steps. In practice this difference does not have a significant effect on training as we will show in the next section.

\subsection{Proof of Concept: Lung Adenocarcinoma EGFR Mutation Prediction}
In this section we describe experiments performed on real-world clinical data where we applied the proposed method to the prediction of EGFR mutations on a large clinical dataset described in~\cite{campanella_hampe-based_2022}. The dataset consists of 2,449 digitized slides from 2,056 lung adenocarcinoma patients collected at *Inst* paired with EGFR mutational status derived from the IMPACT sequencing panel~\cite{cheng_comprehensive_2017}. We used the same Monte Carlo Cross Validation scheme as in~\cite{campanella_hampe-based_2022} where 1,951 slides were used for training and 498 for validation with 20 randomly generated splits following a Monte Carlo Cross-Validation (MCCV) scheme. At each epoch, 50\% of the slides were selected for training to speed up the experiments. For each slide, $K$ tiles were sampled at random. If a slide contained less than $K$ tiles, sampling with replacement was performed. Tiles were loaded on the fly using the \verb|cucim| library from NVIDIA, no pre-tiling of the slides was performed. We jointly trained a ResNet50 and GMA with increasing number of $K$ tiles per slide. With $K \in \{50,100,350,700\}$ we used a single GPU, while with $K \in \{700,2100,4900\}$ we used our parallel implementation with up to 8 GPUs. For each value of $K$ in the multi-GPU setting, we compared training with full precision and AMP. Experiments were run on a cluster of H100 GPUs.

\begin{figure}[h]
\includegraphics[width=\textwidth]{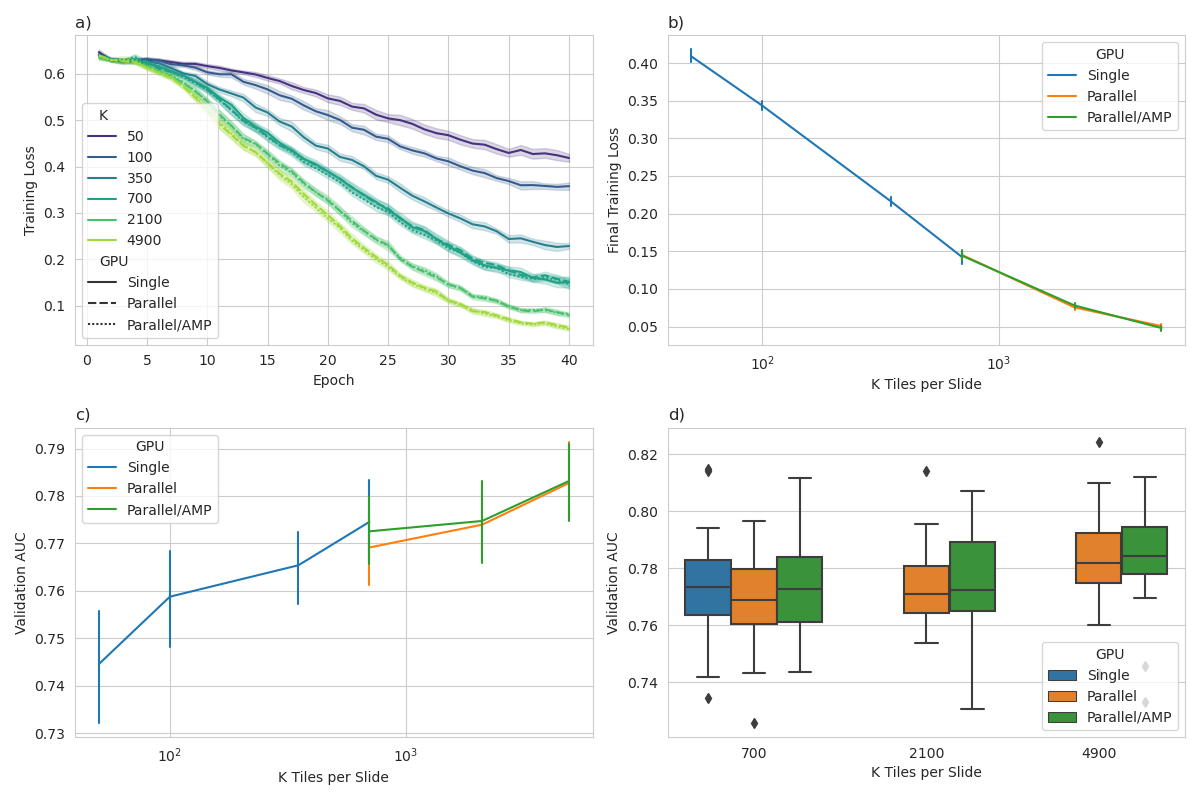}
%\caption{EGFR Mutation Prediction in LUAD Experiments. a) Training loss convergence curves stratified by K tiles per slide and GPU parallelization strategy. Each curve summarizes the 20 MCCV runs. The shaded area is calculated by bootstrapping with 95\% confidence intervals (CI). b) Final Training loss stratified by GPU parallelization strategy in relation to K tiles per slide. Each point is the average over the 20 MCCV runs, and the error bar is estimated via bootstrapping using 95\% CI. c) Validation AUC stratified by GPU parallelization strategy in relation to K tiles per slide. Each point is the average over the 20 MCCV runs, and the error bar is the 95\% CI average estimate calculated via bootstrapping. d) Comparison of validation AUCs between GPU parallelization strategies. Each boxplot represents the distribution of validation AUCs of the 20 MCCV runs.} \label{fig4}
\caption{EGFR Mutation Prediction in LUAD. Each data point summarizes the 20 MCCV runs. a) Training loss convergence curves stratified by $K$ tiles per slide and GPU parallelization strategy. The shaded area is calculated by bootstrapping with 95\% confidence intervals (CI). b) Final Training loss stratified by GPU parallelization strategy in relation to $K$ tiles per slide. The error bar is estimated via bootstrapping using 95\% CI. c) Validation AUC stratified by GPU parallelization strategy in relation to $K$ tiles per slide. The error bar is the 95\% CI average estimate calculated via bootstrapping. d) Comparison of validation AUCs between GPU parallelization strategies.} \label{fig4}
\end{figure}

In Fig.~\ref{fig4} we present the results of these experiments. It can be observed that the training loss decreases as $K$ increases. In concordance with our gradient equivalence experiments, there is no significant difference in loss when comparing single and parallel runs (see $K=700$ in Fig.~\ref{fig4}a,b). Full precision and AMP runs are also comparable in terms of training loss. In Fig.~\ref{fig4}b we plotted the final training loss against $K$ tiles and observed an exponential decay of the loss as $K$ increases.
In terms of validation AUC, we recorded the best validation AUC for each MCCV split. In Fig.~\ref{fig4}c we plot the validation AUC stratified by GPU parallelization strategy in relation to the values of $K$ and observe that the estimate of average AUC increases with larger $K$ values. It is likely that a further decrease in loss and increase in validation AUC could be observed for larger $K$ values. Interestingly, full and half precision result in similar performance in terms of validation AUC (Fig.~\ref{fig4}d).
In summary, we determined that increasing $K$ results in lower training loss and higher validation AUC and confirmed that the various GPU strategies are equivalent.

In Table~\ref{tab1} we compare these results with previous published and unpublished results on this same dataset using a variety of training strategies: i) training in a supervised manner on all tissue tiles with slide-level targets as in Coudray et. al.~\cite{coudray_classification_2018} or ii) only from tumor tiles, iii) using max-pool MIL as in Campanella et. al.~\cite{campanella_clinical-grade_2019}, iv) training a GMA aggregator based on the features learned in strategy ii, v) training a transMIL~\cite{shao_transmil_2021} aggregator based on the features learned in strategy ii, vi) training a GMA aggregator using an ImageNet pre-trained truncated ResNet50 as popularized by~\cite{lu_data-efficient_2021}, vii-viii) training a GMA aggregator using DINO trained ViT models described in~\cite{campanella_computational_2023}. Some of these strategies require significantly more effort and resources. The best performing (strategy iv) requires training a tumor segmentation model, then training a tile encoder model on tumor tiles, and finally training a slide-aggregation model. The next best, strategies vii and viii require access to hundreds of thousands of slides and thousands of GPU hours to train a SSL tile encoder before training a slide encoder. Yet, compared to previous strategies, the proposed method’s performance is on par or better while processing a relatively low number of tiles per slide. %It is interesting to note that while it is possible to process an entire slide with enough GPUs, in many cases it may be enough to use a smaller value of $K$.

\begin{table}
\caption{Comparison of performance on the EGFR prediction task of popular training strategies from published and unpublished results. For the proposed method, we show the results for various values of K and GPU parallelization strategies. S stands for single GPU, P stands for parallel GPUs, H stands for parallel GPUs with half precision. $\dagger$~\cite{campanella_hampe-based_2022}; $\ddagger$~\cite{campanella_computational_2023}; $\ast$ Unpublished}\label{tab1}
\begin{tabular}{lllll}
\toprule
Algorithm & Pre-trained Encoder & Tiles & AUC Avg (Std) & Publication\\
\hline
Encoder training:\\
\hline
Supervised & ResNet50 ImageNet & All & 71.0 (4.7) & $\dagger$\\
Supervised & ResNet50 ImageNet & Tumor & 74.6 (5.3) & $\dagger$\\
Max-pool MIL & ResNet50 ImageNet & All & 68.0 (2.6) & $\ast$\\
\hline
Aggregator training:\\
\hline
GMA & Supervised Tumor & All & 78.8 (2.8) & $\dagger$\\
TransMIL & Supervised Tumor & All & 73.9 (4.2) & $\dagger$\\
GMA & tResNet50 ImageNet & All & 64.9 (3.1) & $\ddagger$\\
GMA & DINO ViT-small 423k & All & 75.3 (3.0) & $\ddagger$\\
GMA & DINO ViT-base 423k & All & 76.6 (2.4) & $\ast$\\
\hline
Proposed:\\
\hline
$K=50$ & ResNet50 ImageNet & All & 74.4 (2.9) & \\
$K=100$ & ResNet50 ImageNet & All & 75.9 (2.5) & \\
$K=350$ & ResNet50 ImageNet & All & 76.5 (1.7) & \\
$K=700$ (S) & ResNet50 ImageNet & All & 77.5 (2.0) & \\
$K=700$ (P) & ResNet50 ImageNet & All & 76.9 (2.9) & \\
$K=700$ (H) & ResNet50 ImageNet & All & 77.3 (1.7) & \\
$K=2100$ (P) & ResNet50 ImageNet & All & 77.4 (1.5) & \\
$K=2100$ (H) & ResNet50 ImageNet & All & 77.5 (2.1) & \\
$K=4900$ (P) & ResNet50 ImageNet & All & 78.3 (1.9) & \\
$K=4900$ (H) & ResNet50 ImageNet & All & 78.3 (1.9) & \\
\bottomrule
\end{tabular}
\end{table}

\subsection{Running on Whole Slides: Breast Cancer Detection}
\begin{figure}[h]
\includegraphics[width=\textwidth]{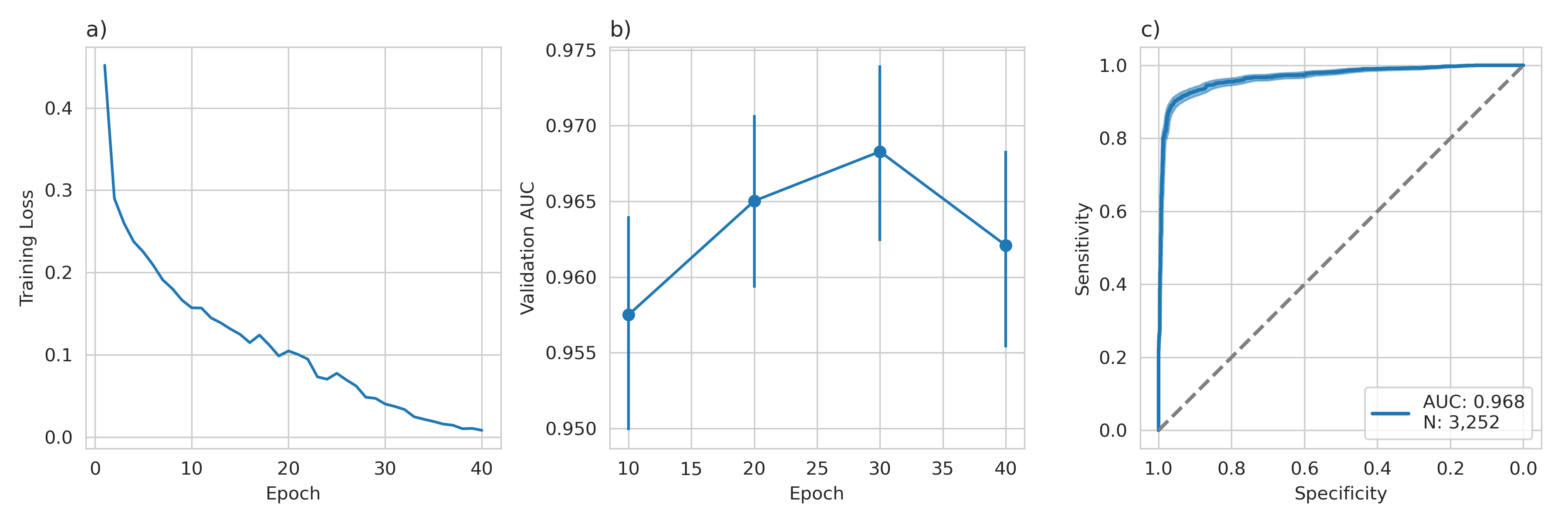}
\caption{Breast Cancer Detection Experiment. a) Training loss convergence. b) Validation AUC convergence. Bars represent bootstrapped 95\% confidence interval. c) ROC curve for the best validation result. The shaded region represents the bootstrapped 95\% confidence interval.} \label{fig5}
\end{figure}

In this section we describe how we applied the proposed system to a cancer detection task and trained a model end-to-end on the entire slide at 20x magnification. From our *Inst*’s research slide archive we queried all scanned and de-identified H\&E breast slides from the beginning of the scanning initiative until September 2023. On these cases, we automatically extracted cancer status (benign vs cancer) at the specimen level from the pathology laboratory information systems (LIS). We were able to obtain an automatically curated dataset of 77,768 slides, where 67,654 were negative and 10,114 were positive. We further sampled this cohort to obtain a balanced dataset of 16,302 slides. The data was then divided at the patient level in a training split consisting of 13,050 slides and a validation split consisting of 3,252 slides.
We calculated the maximum number of tissue tiles of size 224 pixels per slide on this cohort. To allow for full slide analysis, we parallelized encoding with a ResNet18 network on 11 GPUs with $K=4,096$ tiles per GPU, enabling training up to 45,056 tiles per slide.
We trained the encoder and a GMA aggregator jointly, end-to-end on 12 H100 80GB GPUs. At each epoch, 50\% of the slides were sampled for training. We used the AdamW~\citep{loshchilov_decoupled_2017} optimizer and a peak learning rate of $5\cdot10^{-5}$. No hyperparameter tuning was performed. In Fig.~\ref{fig5} we show training and validation results of this experiment. We found a maximum validation AUC of 0.968 after 30 epochs, after which we can see signs of overfitting.

\subsection{Fine-Tuning Foundation Models: Lung Adenocarcinoma EGFR Mutation Prediction}
\begin{figure}[h]
\includegraphics[width=\textwidth]{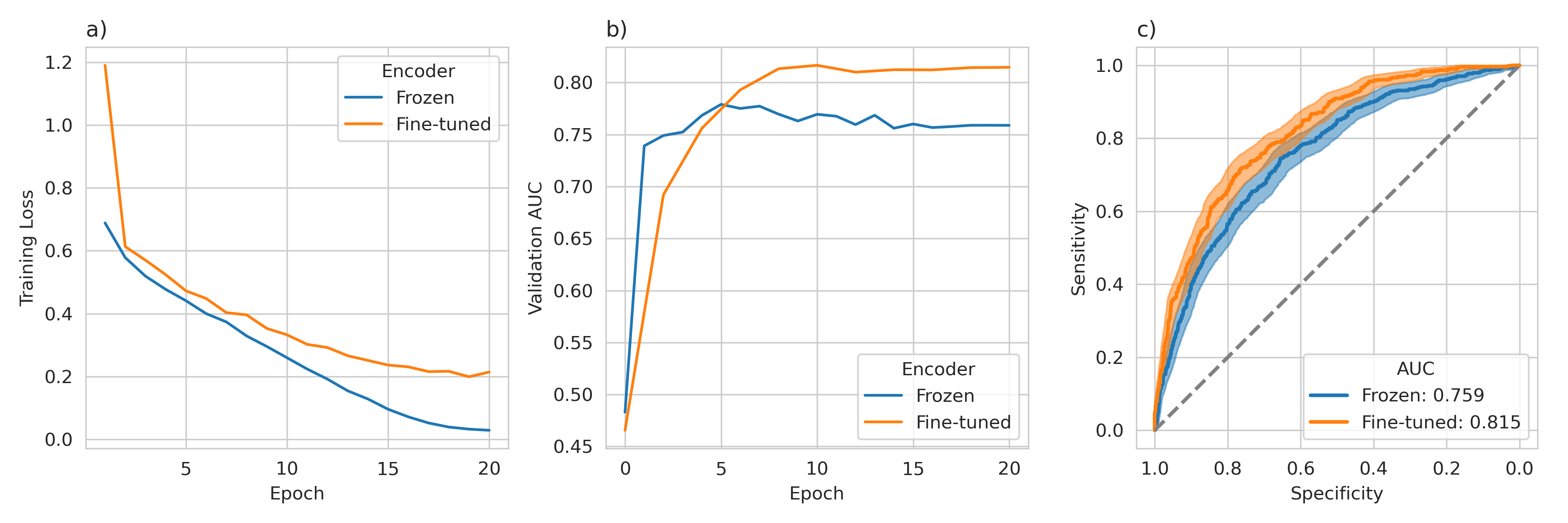}
\caption{Fine-tuning of a Foundation Model to predict EGFR mutations on a large LUAD dataset. A comparison between using a frozen encoder and using the proposed framework to fine-tune it. a) Training loss convergence. b) Validation AUC convergence. c) ROC curves. Shaded regions represent the bootstrapped 95\% confidence interval.} \label{fig6}
\end{figure}

In this section we provide evidence of the benefits of fine-tuning pathology foundation models using the proposed framework. For this experiment, we expanded the dataset described in~\cite{campanella_hampe-based_2022} using the same inclusion criteria. We obtained a total of 6,916 slides from *Inst* with accompanying EGFR mutational status identified via the IMPACT sequencing panel. 5,174 slides were randomly selected for training and 1,742 for testing. The foundation model we used is a ViT-base pre-trained on 423 thousand slides from *Inst* as described in~\cite{campanella_computational_2023}. As a baseline, we train a GMA aggregator model without optimizing the foundation model (frozen encoder). This represents the most common training strategy in computational pathology. We compare this baseline with the training end-to-end of the foundation model and a GMA aggregator using the proposed framework. The GMA-only model was trained on a single GPU using the AdamW optimizer for 20 epochs with a peak learning rate of $10^{-4}$.  The ViT-base with GMA model was trained using mixed precision on 16 H100 GPUs using the AdamW optimizer for 20 epochs with a peak learning rate of $10^{-6}$. For each slide, $15*728$ tiles were sampled at each optimization step. Figure~\ref{fig6} presents the results of this experiment. We observed that the GMA-only model can be trained faster (i.e., with less epochs), but results in inferior performance compared to the model where we allow the optimization of the tile-encoder model. After training, the GMA-only model achieves a validation AUC of 0.76, whereas the fine-tuned model reaches an AUC of 0.82.

\section{Discussion}
Recent advances in computer vision and computational pathology, along with the digitization of pathology data, have enabled the development of decision support systems tailored to a wide variety of clinical tasks, from quality control to survival analysis. Despite the numerous efforts in this space, so far only one system has found its way into clinical practice~\cite{commissioner_fda_2021}.
While diagnostic decision support systems in pathology have met the high standards of clinical application, the same is not true for many other tasks where computational pathology may be of great benefit. These include biomarker prediction, treatment response prediction, and treatment recommendation. While in diagnostic tasks, the signal is generally well defined and recognizable by human experts and is contained within small tissue regions, for more complex tasks the source of the signal, if at all present, is largely unknown. It is possible that local and global spatial arrangements of cells may be informative and methods that can capture both may be more suited for more complex tasks. The majority of work in computational pathology rely on a two-step approach where features are independently learned at a tile-level, and then aggregated at a slide-level. This has proven successful for some tasks, especially cancer detection and classification, but it has not seen the same level of success in others.
By connecting both steps in an end-to-end fashion, it may be possible to obtain a better representation of a pathology slide which could improve performance for certain tasks. Research in this direction is starting to appear (\cite{pinckaers_streaming_2022,wang_when_2023}).

In this paper we present a framework that allows for the analysis of an entire pathology slide at full resolution. Unlike previous efforts, our framework relies on full resolution slides and is modular, allowing encoders and aggregators to be chosen to fit the needs and resources of a particular project.
%In addition, their resizing strategy may be problematic since it can introduce large variations in pixel resolution. This may not be as problematic in TCGA slides, but in real life samples, slide dimensions can vary widely depending on the amount of tissue on the slide.
%Our proposed strategy parallelizes encoding to multiple GPUs and uses customized GPU communications to connect gradients from the slide-level aggregator back to the encoder.
We applied the proposed architecture to a variety of use cases and tasks, demonstrating its flexibility and wide applicability.
%We explored training: with various degrees of sub-sampling, using ImageNet pre-trained models, on entire slides at once, and in the transfer learning setting by fine-tuning modality-specific foundation models. 
Importantly, we strived to provide evidence of clinical deployment performance by leveraging datasets generated during the standard clinical workflow.
We applied the framework to the clinically relevant task of predicting EGFR mutations in lung adenocarcinoma patients where we showed that our end-to-end approach superior to all previous training strategies, especially in the fine-tuning setting. Further, we proved the feasibility of training end-to-end on entire slides at full resolution by applying the proposed framework to a breast cancer detection task.
While the proposed method can be scaled up to analyze an entire slide end-to-end, it is also amenable to more modest set-ups by simply sampling fewer tiles per slide. In our experiments we have shown that this strategy can be effective. Our proposed method can be trained from scratch on a specific task, or fine-tuned from a pre-trained encoder. 

\textbf{Limitations.} The main limitation is related to computational expense. While it is possible to analyze entire slides, it requires significant resources, rendering this approach less appealing for training in resource constrained settings. Despite that, we observed that sub-sampling tile in a slide often provide good performance in the tasks analyzed so far. In addition, experiments shown here focused on the use of clinical datasets, making comparisons with other methods more challenging. More studies will be needed to assess the benefits of this method more broadly.

In this paper we have explored training and fine-tuning models for specific downstream tasks. Yet, approaches such as the one presented here open the way for novel strategies to train foundation models in pathology by learning better tile-encoders and pathology representations directly from slide-level signals. Current strategies for foundation model training rely on applying SSL algorithms to tiles extracted from pathology slides. While these SSL trained encoders are an improvement over natural images pre-trained networks, further improvements could be obtained by leveraging supervised or self-supervised tasks at the slide level.
%we argue that the best use of this framework will be in the training of pathology foundation models based on large-scale clinical datasets. This work opens the possibility to train better encoders based on slide-level targets. Current strategies rely on applying SSL algorithms to tiles extracted from pathology slides. While these SSL trained encoders are an improvement over natural images pre-trained networks, we expect further improvements by directly supervising training on pathology relevant tasks. 
In future experiments we plan to investigate the use of the proposed framework to train a large foundation model from scratch or to align SSL-trained foundation models to specific pathology tasks, based on slide-level signals automatically extracted from the pathology LIS.

\begin{ack}
This work is supported in part through the use of research platform AI-Ready Mount Sinai (AIR.MS), the expertise provided by the team at the Hasso Plattner Institute for Digital Health at Mount Sinai (HPI.MS), and through the computational and data resources and staff expertise provided by Scientific Computing and Data at the Icahn School of Medicine at Mount Sinai and supported by the Clinical and Translational Science Awards (CTSA) grant UL1TR004419 from the National Center for Advancing Translational Sciences.
%This study has been approved by the Mount Sinai Institutional Review Board (IRB #19-0051).
\end{ack}

\section{References}

\bibliographystyle{plainnat} % Choose your preferred style
\bibliography{bibliography} % The file containing your references
%%%%%%%%%%%%%%%%%%%%%%%%%%%%%%%%%%%%%%%%%%%%%%%%%%%%%%%%%%%%

\appendix

\section{Appendix / supplemental material}
\begin{lstlisting}[language=Python, caption=Pytorch pseduo-code implementation, label=code]
# Initialize GPU processes
# Initialize data loader and models
# Initialize DDP group for encoder model
DDPgroup = dist.new_group(ranks=[1, ..., N])
encoder_model = DDP(encoder_model, group=DDPgroup)
for epoch in range(E):
    for i, batch in enumerate(loader):
        # Note that each optimization step is done for one slide
        # Data loader and distributed sampler are in charge of \
        feeding the right data to each process
        
        # Forward pass on encoder
        if rank != 0:
            features = encoder(batch)
        
        # Gather features
        with torch.no_grad():
            if rank == 0:
                 dist.gather(features, allfeatures,\
                 dst=0, group=None)
        
        # Forward/Backward pass on aggregator
        if rank == 0:
            # Record gradients on features
            allfeatures.requires_grad_()
            # Forward pass aggregator
            output = aggregator(allfeatures)
            aggregator_loss = criterion(output, label)
            # Backward pass aggregator
            aggregator_loss.backward()    
            # Get feature gradients
            grads = allfeatures.grad.detach()
        
        # Scatter feature gradients
        with torch.no_grad():
            dist.scatter(grads_recv, grads, src=0, group=None)
        
        # Backward pass encoder
        # Generate loss on DDP group
        # Loss has to be scaled by number of DDP processes
        if rank != 0:
            encoder_loss = \
            pseudo_loss(features, grads_recv) * (world_size-1)
            encoder_loss.backward()
\end{lstlisting}

\end{document}